\documentclass[oldversion]{aa}
\usepackage{txfonts}
\usepackage{natbib}
\bibpunct{(}{)}{;}{a}{}{,}
\usepackage{longtable}
\usepackage[dvips]{graphicx}
\newcommand{\chandra}{\textit{Chandra}}
\newcommand{\ms}{$M_{\star}$}
\newcommand{\sn}{$S_{\rm N}$}
\newcommand{\re}{$r_{\rm e}$}
\newcommand{\ks}{$K_{\rm S}$}

\newcommand{\ninx}{$N^{\rm in}_{\rm X}$}
\newcommand{\nincxb}{$N^{\rm in}_{\rm CXB}$}
\newcommand{\ninks}{$N^{\rm in}_{\rm KS}$}
\newcommand{\noutx}{$N^{\rm out}_{\rm X}$}
\newcommand{\noutcxb}{$N^{\rm out}_{\rm CXB}$}
\newcommand{\noutks}{$N^{\rm out}_{\rm KS}$}
\newcommand{\inreg}{(0.2 -- 3)$r_{\rm e}$}
\newcommand{\outreg}{(4 -- 10)$r_{\rm e}$}

\begin{document}

\title{LMXB populations in galaxy outskirts: globular clusters and supernova kicks}
\author{Zhongli Zhang \inst{1} \and Marat Gilfanov \inst{1,2} \and {\'A}kos Bogd{\'a}n \inst{3} }
\institute{Max-Planck Institut f\"{u}r Astrophysik, Karl-Schwarzschild-Stra\ss e 1, D-85741 Garching, Germany 
\and Space Research Institute, Russian Academy of Sciences, Profsuyuznaya 84/32, 117997 Moscow, Russia 
\and Smithsonian Astrophysical Observatory, 60 Garden Street, Cambridge, MA 02138, USA, Einstein Fellow
\\
email:[zzhang;gilfanov]@mpa-garching.mpg.de}
\titlerunning{LMXB populations in galaxy outskirts}
\date{Received ... / Accepted ...}

\abstract{ 
For the first time, we have systematically explored the population of discrete X-ray sources in the outskirts of early-type galaxies. Based on a broad sample of 20 galaxies observed with \chandra\ we detected overdensity of X-ray sources in their outskirts. The overdensity appears as halos of resolved sources around the galaxies. These halos are broader than the stellar light, extending out to at least $\sim 10 r_e$ ($r_e$ is the effective radius). These halos are composed of sources fainter than $\sim5\times10^{38}$ erg/s, whereas the more luminous sources appear to follow the distribution of the stellar light, suggesting that the excess source population consists of neutron star binaries. Dividing the galaxy sample into four groups according to their stellar mass and  specific frequency of globular clusters, we find that the extended halos are present in all groups except for the low-mass galaxies with low globular cluster content. We propose that the extended halos may be comprised of two independent components, low-mass X-ray binaries (LMXBs) located in globular clusters (GCs), which are known to have a wider distribution than the stellar light, and neutron star (NS) LMXBs kicked out of the main body of the parent galaxy by supernova explosions. The available deep optical and X-ray data of NGC 4365 support this conclusion. For this galaxy we identified $60.1\pm10.8$ excess sources in the \outreg\ region of which $\sim40$\% are located in GCs, whereas $\sim60$\% are field LMXBs. We interpret the latter as kicked NS LMXBs. We discuss the implications of these results for the natal kick distributions of black holes and neutron stars. 
}
\maketitle


\begin{table*}
\begin{center}
\caption{The galaxy sample.}
\label{tab:sample}
\begin{tabular}{lcccccccccccc}
\hline
\hline
Galaxy&   $D$  & Scale        & $r_{\rm e}$ & b/a   & PA     & $L_{\rm K}$		  & $M_{\star}/L_{\rm K}$	   &\ms 		 & \sn  & Age	& Exp & $L^{\rm in}_{\rm lim}$/$L^{\rm out}_{\rm lim}$ \\
      &   (Mpc)& (pc/arcsec)  & (arcsec)    &	    &(degree)&($10^{10}L_{\rm K,\odot}$)  & ($M_{\odot}/L_{\rm K,\odot}$)  & ($10^{10}M_{\odot}$)&	& (Gyr) & (ks)& ($10^{37}$ erg/s) \\
      &  (1)   &  (2)	      & (3)	    & (4)   & (5)    &    (6)			  &	  (7)			   & (8)		 & (9)  &  (10) & (11)&  (12)      \\ 
\hline  
\\
N720  &  27.7  &   134    & 25.2	& 0.55  & -40.0  & 21.50 & 0.86 & 18.49  & 1.01$^a$ & 3.4$^{aa}$  & 138.8 & 5.6/4.1   \\
N821  &  24.1  &   117    & 19.6	& 0.62  &  30.0  & 9.12  & 0.82 & 7.48   & 1.14$^b$ & 5.2$^{bb}$  & 212.9 & 2.0/2.0   \\
N1052 &  19.4  &   94     & 18.6	& 0.70  & -60.0  & 8.94  & 0.80 & 7.15   & 1.59$^c$ & 14.5$^{cc}$ &  59.2 & 3.9/2.5   \\
N1380 &  17.6  &   85     & 31.8	& 0.44  &  7.0   & 12.57 & 0.81 & 10.18  & 1.81$^d$ & 4.4$^{cc}$  &  41.6 & 3.5/2.9   \\
N1404 &  21.0  &   102    & 18.6	& 0.90  & -17.0  & 18.73 & 0.85 & 15.92  & 1.69$^e$ & 5.9$^{aa}$  & 114.5 & 5.7/2.9   \\
N3115 &  9.7   &   47     & 33.6	& 0.39  &  45.0  & 9.43  & 0.83 & 7.83   & 2.22$^f$ & 8.4$^{bb}$  & 153.2 & 0.45/0.42 \\
N3379 &  10.6  &   51     & 28.5	& 0.85  &  67.0  & 7.92  & 0.83 & 6.57   & 1.20$^g$ & 8.2$^{bb}$  & 337.0 & 0.33/0.33 \\
N3585 &  20.0  &   97     & 29.9	& 0.63  &  -75.0 & 18.92 & 0.77 & 14.57  & 0.50$^h$ & 3.1$^{aa}$  &  94.7 & 2.6/2.7   \\
N3923 &  22.9  &   111    & 40.5	& 0.64  & 47.0   & 29.90 & 0.82 & 24.52  & 3.57$^i$ & 3.3$^{dd}$  & 102.1 & 3.6/3.4   \\
N4125 &  23.9  &   116    & 31.4	& 0.63  & 82.0   & 23.49 & 0.80 & 18.79  & 1.30$^h$ & 5.0$^{ee}$  &  64.2 & 4.4/4.1   \\
N4278 &  16.1  &   78     & 18.3	& 0.90  & 35.0   & 7.87  & 0.78 & 6.14   & 5.35$^f$ & 12.5$^{bb}$ & 470.8 & 0.62/0.60 \\
N4365 &  20.4  &   99     & 38.1	& 0.74  & 45.0   & 20.86 & 0.85 & 17.73  & 3.95$^j$ & 7.9$^{bb}$  & 195.8 & 1.5/1.5   \\
N4374 &  18.4  &   89     & 33.5	& 0.92  & -57.0  & 24.94 & 0.83 & 20.70  & 5.39$^k$ & 9.8$^{cc}$  & 115.5 & 1.7/2.2   \\
N4382 &  18.5  &   90     & 54.9	& 0.67  & 12.0   & 27.06 & 0.76 & 20.57  & 1.43$^k$ & 1.6$^{aa}$  &  39.7 & 2.9/4.1   \\
N4472 &  16.3  &   79     & 56.1	& 0.81  & -17.0  & 41.88 & 0.85 & 35.60  & 6.61$^k$ & 9.6$^{bb}$  &  89.6 & 3.2/3.7   \\
N4552 &  15.3  &   74     & 22.8	& 0.94  & -30.0  & 10.82 & 0.83 & 8.98   & 2.99$^k$ & 6.0$^{cc}$  &  54.4 & 1.8/1.8   \\
N4636 &  14.7  &   71     & 56.2	& 0.84  & -37.0  & 13.24 & 0.81 & 10.72  & 12.38$^l$& 13.5$^{cc}$ & 209.8 & 1.4/1.4   \\
N4649 &  16.8  &   81     & 42.1	& 0.81  & -72.0  & 32.44 & 0.85 & 27.57  & 5.32$^k$ & 16.9$^{bb}$ & 108.0 & 2.3/2.3   \\
N4697 &  11.7  &   57     & 39.5	& 0.63  & 67.0   & 8.82  & 0.77 & 6.79   & 3.78$^m$ & 10.0$^{cc}$ & 193.0 & 0.59/0.59 \\
N5866 &  15.3  &   74     & 28.5	& 0.42  & -57.0  & 9.47  & 0.72 & 6.82   & 1.69$^n$ & 1.8$^{aa}$  &  33.7 & 2.1/1.9   \\
\hline
\end{tabular}
\end{center}
{\bf Notes.} (1) -- Galaxy distance derived by \citet{Tonry2001}. (2) -- Scale conversion. (3), (4), and (5) -- \ks-band half-light radius, axis-ratio, and position angle from the 2MASS Large Galaxy Atlas \citep{Jarrett2003}. (6) -- Total \ks-band luminosity calculated from the total apparent \ks-band magnitude. (7) -- \ks-band mass-to-light ratios derived from \citet{Bell2001}, with $B-V$ colors from RC3 catalog \citep{Dev1991}. (8) -- Total stellar mass calculated from \ks-band total luminosity and the \ks-band mass-to-light ratio. (9) -- Globular cluster specific frequencies derived following \citet{Zhang2012}. References for the number of observed globular clusters in each galaxy are -- 
 $^a$\citet{Kissler1996}; $^b$\citet{Spitler2008};
 $^c$\citet{Forbes2001}; $^d$\citet{Kissler1997};
 $^e$\citet{Forbes1998}; $^f$\citet{Harris1991};
 $^g$\citet{Rhode2004}; $^h$\citet{Humphrey2009}; $^i$\citet{Sikkema2006};
 $^j$\citet{Forbes1996}; $^k$\citet{Peng2008}; $^l$\citet{Dirsch2005};
 $^m$\citet{Dirsch1996}; $^n$\citet{Cantiello2007}. (10) -- Stellar age of the galaxy. References -- $^{aa}$\citet{Terlevich2002}; $^{bb}$\citet{Sanchez2006}; $^{cc}$\citet{Annibali2007}; $^{dd}$\citet{Thomas2005}; $^{ee}$\citet{Schweizer1992}. (11) -- Total exposure time of the combined \chandra\ observations. (12) -- Limiting source detection sensitivity in the  $0.5-8$ keV band in the \inreg\ ($L^{\rm in}_{\rm lim}$) and outer \outreg\ ($L^{\rm out}_{\rm lim}$) regions.
\end{table*}

\section{Introduction}
\label{sec:introduction}

Low-mass X-ray binaries (LMXBs) consist of a compact object, either a
black hole (BH) or a neutron star (NS), and a low-mass donor star
($\lesssim1 \ \rm{M_{\odot}}$) that transfers mass via Roche-lobe
overflow. The infalling matter is heated to X-ray temperatures and
releases X-ray luminosities in the range of $10^{35}-10^{39}$ erg/s. 
Because of their luminous nature, LMXBs add a major contribution to the total 
X-ray emission of galaxies; moreover, they determine the X-ray appearance of 
relatively gas-poor early-type galaxies \citep[e.g.,][]{Fabbiano1989}.
The populations of LMXBs have been extensively studied 
in early-type galaxies with \chandra\ observations \citep[e.g.,][see \citet{Fabbiano2006} for an extensive review]{Kundu2002, Irwin2003,Sarazin2003,Kim2006, Bogdan2011, Zhang2011, Zhang2012}. A simple picture of the correlation of their properties with the parameters of the host galaxy emerged. In particular, it has been established that LMXB population scales with the stellar mass of the host galaxy \citep{Kim2004,Gilfanov2004, Colbert2004}. Moreover, the luminosity distribution of LMXBs in nearby galaxies can be described by a universal luminosity function \citep{Kim2004,Gilfanov2004}.

Investigating the spatial distribution of LMXBs in  NGC 4472, \citet{Kundu2002}  came to the conclusion that it follows more closely the distribution of globular clusters rather than that of the stellar light. Other authors, to the contrary, did not find any significant deviations of the LMXB distribution from the stellar light, for example  in M84 \citep{Finoguenov2002}, NGC 1316 \citep{Kim2003}, and NGC 1332 \citep{Humphrey2004}. Based on large samples of early-type galaxies, \citet{Gilfanov2004} and \citet{Kim2006} also concluded that bright LMXBs follow stellar light distribution, except for the very inner regions of galaxies \citep{Voss2007}.
In general, this point of view seemed to have become dominant, although not without a hint of controversy. Now, in the course of more than ten years of operation of \chandra, a large number of 
early-type galaxies have been observed to a fairly deep sensitivity limit ($L_{\rm X} \sim 10^{37}$ erg/s), 
allowing us to probe LMXB populations in unprecedented details.

Such a detailed study of LMXB populations has been performed for the Sombrero
galaxy (M104) by \citet{Li2010}. Their results appear to challenge again the
empirical picture that the LMXB populations closely follow the
stellar light distribution at all galactocentric radii. As opposed to the
expectations, the outskirts of Sombrero exhibit a significant X-ray source
excess. In the halo of this galaxy $101$ sources were detected, whereas
the expected number of cosmic X-ray background (CXB) sources is $52\pm11$, implying a
$\sim4.4\sigma$ excess. Although \citet{Li2010} did not identify the origins 
of the excess X-ray sources, they considered that either supernova kicked 
binary systems and/or LMXBs associated with GCs could be responsible.

Globular clusters have a bimodal color distribution \citep{Forbes1997}. The red
(metal-rich) population follows the stellar light distribution, whereas
the blue (metal-poor) population has a wider distribution and traces the
more extended dark matter halos \citep[e.g.,][]{Bassino2006}. Since a small
percentage of blue GCs host LMXBs \citep{Kundu2002,Jordan2004}, it is likely that
LMXBs associated with blue GCs will have a wider distribution than the stellar 
light. This effect could result in an extended population 
of LMXBs in the outskirts of galaxies. Obviously, the importance of GC-LMXBs is 
expected to be more prominent in galaxies with a higher globular cluster specific 
frequency (\sn).

Neutron stars receive kicks (natal kicks) when they are formed in the core collapse 
supernova explosions. The amplitude of the natal kick velocities can be determined; 
for example, from proper motion of pulsars, the mean birth speed for young pulsars is $\sim 400$ km/s 
with the fastest neutron stars moving with a velocity in excess of $\sim 10^3$ km/s  \citep{Hobbs2005}. 
Although only a relatively small fraction of binary systems can survive large natal 
kicks \citep{Brandt1995}, those that do survive can travel to large distances producing a distribution 
much wider than that of the stars. The effect of the natal kicks on binaries should be stronger in 
lower mass galaxies where kick velocities may exceed the escape velocity in the gravitational potential 
of the parent galaxy, and binary systems may leave the galaxy or be displaced to very large distances 
from its main body where the binaries were formed.

In this paper, we aim to comprehensively explore the LMXB populations in a large sample 
of galaxy outskirts with the limiting sensitivity of $\sim 10^{37}\ \rm{erg \ s^{-1}}$. We 
will address two major points. First, we investigate whether the existence of excess 
LMXBs is ubiquitous in the outskirts of early-type galaxies. Second, we study the origin 
of the excess sources with a particular focus on GC-LMXBs and supernova kicked X-ray binaries. 
To achieve our goals, we use a sample of 20 early-type galaxies with deep \chandra\ 
observations. Since our galaxy sample covers a broad range in both stellar mass and globular 
cluster specific frequency, we are able to perform a systematic study and address the 
importance of GC-LMXBs and supernova kicked LMXBs.

The paper is structured as follows. In Sect.~\ref{sec:sample} we introduce the analyzed sample. 
In Sect.~\ref{sec:data} the X-ray and near-infrared data analysis is described. In Sect.~\ref{sec:excess} and Sect. 5,  
we discuss the radial distributions and the X-ray luminosity distributions  of the detected X-ray sources.  In Sect.~\ref{sec:n4365} 
we describe a case study for NGC 4365. Our results are discussed in Sect.~\ref{sec:discussion} and 
we give our conclusions in Sect.~\ref{sec:conclusion}.

\section{The analyzed sample}
\label{sec:sample}

In the present work we explore the 
LMXB populations in the outskirts of early-type (E/S0) galaxies. To achieve this goal and obtain 
statistically significant conclusions, a broad sample of galaxies must be analyzed. 
In \citet{Zhang2012} we carefully built a sample of 20 galaxies, which also  well suits the proposed 
goals of the present analysis for the following four reasons. Firstly, the sample consists of galaxies 
within the distance range of $9.7-27.7$ Mpc. Therefore, adequate ($\sim5\times 10^{37}\ \rm{erg \ s^{-1}}$) 
source detection sensitivities can be achieved  with moderately deep ($<150$ ks) \chandra\
exposures, and the sample galaxies and their outskirts fit well within the ACIS field-of-view. 
Secondly, the sample includes both GC-rich and GC-poor galaxies (\sn$=0.50-12.38$). Thirdly, the sample 
covers a relatively broad range of stellar masses ($M_{\star} =(6.1-35.6)\times10^{10} \ M_{\odot}$), 
implying vastly different numbers of field LMXBs, as well as different dark matter halo masses. Finally, 
the selected galaxies do not exhibit ongoing or recent star-formation, hence their stellar content 
is fairly homogeneous and the population of X-ray binaries is not polluted by high-mass X-ray binaries.
The physical properties of the sample galaxies are listed in Table~\ref{tab:sample}.

\section{Data analysis}
\label{sec:data}

\subsection{\chandra\ data}
\label{sec:chandra}

The analyzed \chandra\ observations and the main steps of the data analysis agree with those outlined 
in \citet{Zhang2012}. The data was reduced using standard CIAO threads (CIAO version 4.2; CALDB version 4.2.1).
To detect point sources, we applied the CIAO \textit{wavdetect} task with parameters taken from \citet{Voss2006,Voss2007}. 
Namely, we set the threshold parameter (\textit{sigthresh}) to $10^{-6}$, which implies one false detection per $10^6$ pixels 
or one ACIS-S CCD. The source detection was performed in the $0.5-8$ keV band. To increase the source detection 
sensitivity, we did not exclude high background periods since they are outweighed by the increased exposure 
time. We produced exposure maps in the $0.5-8$ keV band, assuming a single power-law model with $\Gamma=1.7$ and 
Galactic absorption. Several galaxies in our sample have been observed in multiple \chandra\ pointings; therefore, 
we corrected the offsets following \citet{Voss2007} using the CIAO \textit{reproject\_events} tool. After correcting the offsets, the images were combined and re-analyzed. 

To compute the source net counts, we employed circular apertures centered on the central coordinates of each source. 
The radius of the aperture was defined as a circle, which includes $85\%$ of the point spread function (PSF) value. 
For each source, the PSF was determined using the CIAO \textit{mkpsf} tool. To account for the background components, 
we applied circular regions with three times the radius of the source regions, while the overlapping regions of adjacent 
sources were excluded. The source net counts were derived following \citet{Voss2007}. The observed net counts were 
converted to $0.5-8$ keV band unabsorbed luminosities assuming a power-law spectrum ($\Gamma=1.7$) with Galactic 
absorption.

\begin{table}
\begin{center}
\caption{The statistics of X-ray point sources.}
\label{tab:stat}
\begin{tabular}{ccccccc}
\hline
\hline
Galaxy&  \ninx & \nincxb & \ninks & \noutx & \noutcxb & \noutks \\
      &  (1)   & (2)	 & (3)    & (4)    & (5)      & (6)	\\     
\hline 
\\
N720  &   50   &   1.6   &  31.7  &   58   &  16.8    &   7.9	\\
N821  &   34   &   1.9   &  43.6  &   23   &  16.5    &  13.9	\\
N1052 &   28   &   0.9   &  18.8  &   19   &   7.6    &   6.2	\\
N1380 &   27   &   1.5   &  31.0  &   14   &  12.4    &   3.2	\\
N1404 &   18   &   0.9   &  19.6  &   33   &  12.1    &   8.1	\\
N3115 &   61   &   2.8   &  86.7  &   54   &  21.6    &  12.9	\\ 
N3379 &   68   &   6.4   &  86.1  &   58   &  45.4    &  12.5	\\ 
N3585 &   49   &   2.9   &  63.6  &   44   &  21.8    &   8.3	\\ 
N3923 &   76   &   4.9   &  67.9  &   64   &  26.9    &   6.5	\\
N4125 &   31   &   2.6   &  38.0  &   33   &  23.1    &   3.8	\\
N4278 &  108   &   3.0   &  60.1  &   94   &  27.6    &  12.8	\\
N4365 &  153   &   7.8   & 117.7  &  116   &  44.6    &  11.3	\\
N4374 &   96   &   5.0   &  77.9  &   65   &  30.6    &  12.4	\\ 
N4382 &   37   &   7.2   &  72.9  &   30   &  18.1    &   2.4	\\
N4472 &  151   &   8.7   & 126.6  &   55   &  17.3    &   9.1	\\
N4552 &   52   &   1.9   &  29.7  &   64   &  18.1    &   8.7	\\ 
N4636 &  116   &  11.5   &  59.0  &   71   &  30.5    &   0.1	\\ 
N4649 &  157   &   4.8   &  83.4  &  100   &  26.8    &   6.3	\\
N4697 &   80   &   6.9   &  74.0  &   69   &  39.4    &   7.7	\\
N5866 &   20   &   1.2   &  29.0  &   13   &  10.4    &   4.0	\\
\hline
Total &  1412  &  84.4   & 1217.3 &  1077  & 467.6    & 158.1	\\ 
\hline
\end{tabular}
\end{center}
{\bf Notes.} (1), (2), and (3) -- Number of detected X-ray 
sources, estimated CXB sources, and estimated LMXBs from \ks-band light in the \inreg \ region. 
(4), (5), and (6) -- The same quantities in the \outreg \ region.
\end{table}

The source detection sensitivity varies throughout the \chandra\ images because of the varying 
level of the diffuse X-ray emission associated with the galaxy, the PSF deterioration at 
large off-axis angles, and the non-uniform exposure of the combined images. For the combined 
images, we derived the source detection sensitivities by  inverting the detection method, using 
the local PSF, background, and exposure \citep{Voss2006}. The incompleteness function, $K(L)$, 
in a certain area was calculated by accumulating the sensitivities of the pixels included in the area, 
weighted by the assumed spatial distribution of sources. Thus, $K(L)$ was computed separately for the 
CXB sources with a flat spatial distribution, and for LMXBs assuming that they follow the \ks-band 
stellar light. We used these functions to calculate the number of predicted CXB sources 
 \citep{Geo2008} and LMXBs (Table~\ref{tab:stat}), and to obtain their
predicted radial source density profiles (Sect.~\ref{sec:radial}). The incompleteness 
function was also used to correct apparent  X-ray luminosity functions (XLFs), and 
to produce the final XLFs presented in Sect.~\ref{sec:xlf}.

The contribution of CXB sources is estimated based on their luminosity function from \citet{Geo2008}. 
To convert their $0.5-10$ keV band $\log N - \log S$ distribution to the $0.5-8$ keV band, we assumed 
a power-law model with $\Gamma=1.4$. Within the central regions of the galaxies, the 
contribution of CXB sources is $\lesssim10\%$ (Table~\ref{tab:stat}). Although in the outskirts 
of the sample galaxies the ratio of predicted CXB sources is significantly higher (Table~\ref{tab:stat}), 
in most galaxies a statistically significant source excess is detected above the predicted CXB level. 
We stress that the excess sources cannot be attributed to the cosmic variance, which can be responsible 
for variations on the $\sim 10-30\%$ level. The accuracy of CXB subtraction is further discussed in 
Sect.~\ref{sec:cxb}. 

\subsection{Near-infrared data analysis}
\label{sec:kband}

To trace the stellar light of the sample galaxies, we relied on the \ks-band (2.16 $\mu$m) images of the 
2MASS Large Galaxy Atlas \citep{Jarrett2003}. The provided images are background subtracted, except for 
NGC 821 and NGC 1052, where we estimated the background level using nearby regions. To avoid pollution by 
bright foreground and background objects, we visually 
identified and removed them. We converted the observed number of \ks-band source counts ($S$) to apparent 
magnitudes using $m_{\rm K} = \rm{KMAGZP} -2.5 \log S$, where KMAGZP is the zero point magnitude for the 
\ks-band image. The magnitudes obtained were used to compute \ks-band luminosities, assuming that 
the absolute \ks-band magnitude of the Sun is $M_{\rm{K,\odot}} = 3.39$ mag. 

Based on the \ks-band luminosities and the \ks-band 
mass-to-light ratios, we derived the stellar mass of the sample galaxies (Table~\ref{tab:sample}). 
The mass-to-light ratios were derived from the $B-V$ color indices \citep[RC3 catalog;][]{Dev1991} 
and the results of galaxy evolution modeling \citep{Bell2001}.

\begin{figure}
\resizebox{\hsize}{!}{\includegraphics[angle=270]{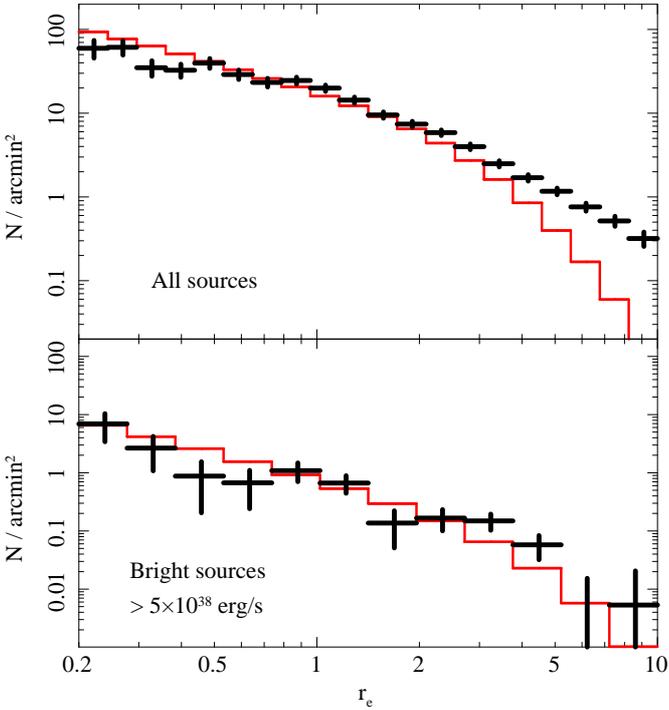}}
\caption{Stacked radial source density profiles of the  CXB subtracted LMXBs in the sample galaxies 
(crosses). The upper and lower panels correspond to all the sources and sources brighter 
than $5\times10^{38}$ erg/s, respectively. The profiles are not corrected for incompleteness. The predicted 
distributions  of LMXBs based on the \ks-band light are plotted with solid histograms.
They take into account the source detection incompleteness, as described in Sect. \ref{sec:chandra}, 
and therefore can be directly compared with observed profile. An excess of LMXBs beyond $\sim$4\re\ is present for 
all sources, but is absent for bright sources.} 
\label{fig:all_bright}
\end{figure}

\section{Excess X-ray sources in galaxy outskirts}
\label{sec:excess}

\subsection{Radial distributions of all galaxies}
\label{sec:radial}

To map the spatial distribution of X-ray sources in and around the sample galaxies, we built 
stacked radial source density profiles (Fig.~\ref{fig:all_bright}). 
The profiles were extracted from concentric ellipses, whose shape and orientations were determined 
by the \ks-band photometry of the galaxies (Table~\ref{tab:sample}). The contribution of CXB 
sources was subtracted as described in Sect.~\ref{sec:chandra}. The observed source 
density profiles are compared with those expected based on the average XLF of 
LMXBs \citep{Zhang2012}. 

In the upper panel of Fig.~\ref{fig:all_bright} we show the source density profile for all X-ray 
sources without employing a luminosity cut. Within the \inreg\  region the observed and predicted radial profiles are 
in fairly good agreement with each other. Instead, in the \outreg\ region a highly 
significant source excess is detected. 
At first sight, this result appears to contradict the source density profile in \citet{Gilfanov2004}, 
who concluded that the distribution of LMXBs closely follows the stellar light. However, at the time of 
writing, only relatively shallow \chandra\ observations of early-type galaxies were available, and so 
\citet{Gilfanov2004} considered only bright X-ray sources. In the lower panel of Fig.~\ref{fig:all_bright} 
we present the radial source density profile including only sources brighter than $5\times10^{38}$ erg/s. 
The observed distribution of bright sources follows the stellar light distribution at all radii, and are in 
good agreement with \citet{Gilfanov2004}. 

Figure \ref{fig:all_bright} clearly shows that there is an excess in the number of faint sources at large 
radii. The origin and characteristics of the faint excess sources are discussed throughout the paper.

\begin{figure*}
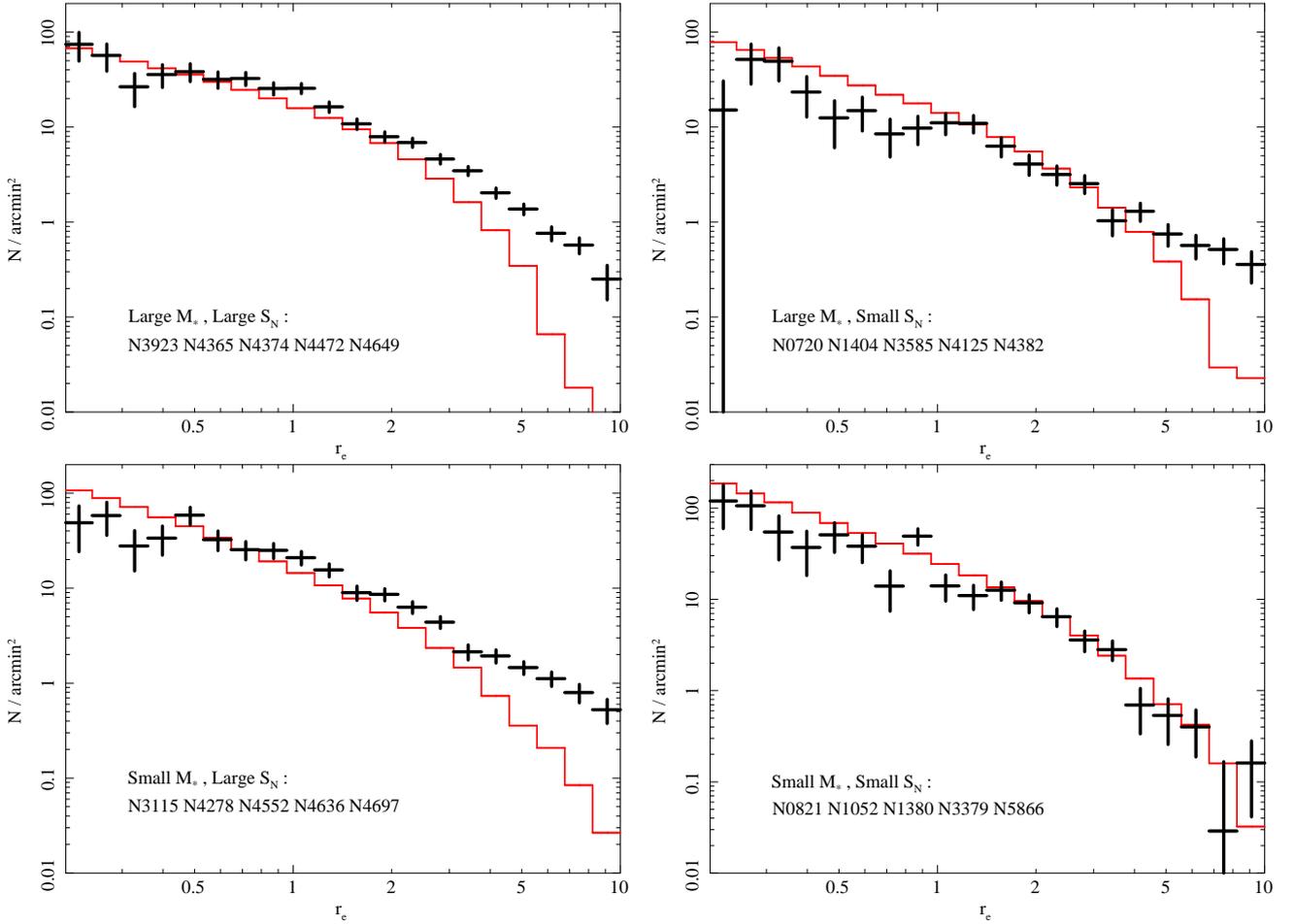

\begin{center}
\resizebox{0.47\hsize}{!}{\includegraphics[angle=270]{lam_lasn.ps}}
\resizebox{0.47\hsize}{!}{\includegraphics[angle=270]{lam_smsn.ps}}\\
\resizebox{0.47\hsize}{!}{\includegraphics[angle=270]{smm_lasn.ps}}
\resizebox{0.47\hsize}{!}{\includegraphics[angle=270]{smm_smsn.ps}}
\caption{Stacked radial source density profiles of the CXB subtracted LMXBs in four groups of galaxies 
with different \ms\ and \sn, marked in the plots. The profiles are not corrected for incompleteness. The predicted 
distributions  of LMXBs based on the \ks-band light are plotted with solid histograms. They take into account the source detection incompleteness, 
as described in Sect. \ref{sec:chandra}. All groups show excess LMXBs in their outskirts, except for galaxies with small \ms\ and small \sn.}
\label{fig:groups}
\end{center}
\end{figure*}

\subsection{Source excess in individual galaxies}
\label{sec:stat}

In Fig.~\ref{fig:all_bright} we demonstrate the existence of an excess population of faint 
sources beyond $\sim4$\re. However, the presented stacked source density profiles were obtained for $20$ galaxies, 
hence it reflects the general behavior of our sample. To study individual galaxies, we derived the 
numbers of detected and predicted X-ray sources in each galaxy separately. Given the relatively low 
number of detected sources in individual systems, we did not attempt to build radial source density 
profiles for each galaxy. Instead, we divided the galaxies into two regions: the inner region was defined 
as an elliptic annulus with \inreg\ radius, whereas the outer region is an elliptic annulus with \outreg. 
We note that the central $0.2$\re\ region was excluded to avoid the possible source confusion.  
The results are summarized in Table~\ref{tab:stat} where we list the numbers of detected sources, predicted LMXBs, 
and estimated CXB sources in the inner and outer regions. As before, the predicted numbers of LMXBs were derived 
using the average LMXB XLF \citep{Zhang2011} and the stellar mass enclosed in the regions. 
The estimated CXB level was derived as described in Sect.~\ref{sec:chandra}.

In the inner region the numbers of detected and predicted X-ray sources agree within $\sim50\%$ for all 
galaxies, except for NGC 4382, NGC 4636, and NGC 4649, where they agree within a factor of $\sim2$. 
In the entire sample, the total number of LMXBs in the inner region after CXB subtraction is $1327.6$, 
whereas the expected number is $1217.3$; these numbers agree within $\sim10\%$. The total stellar mass 
in the \inreg\ region is $2.1\times10^{12} \ \rm{M_{\odot}}$, implying a sample-averaged LMXB specific 
frequency of $6.3\pm0.2$ sources per $10^{10} \ \rm{M_{\odot}}$, in good agreement with \citet{Gilfanov2004} 
and \citet{Zhang2012}. 
In the outer region, $14$ galaxies out of $20$ show a notable X-ray source excess. In most galaxies the number of additional X-ray 
sources exceed the predicted number of LMXBs by a factor of $2.5-6$, and in NGC 4636 and NGC 4649 this 
ratio is even larger. In the outer region we detected $609.4$ X-ray sources after subtracting the CXB level, 
whereas $158.1$ sources are expected, implying an excess of a factor of $\sim3$. The total enclosed stellar mass 
in the outer region is $2.1\times10^{11} \ \rm{M_{\odot}}$, hence the sample-averaged LMXB specific frequency 
is $29.0\pm 1.2$ sources per $10^{10} \ \rm{M_{\odot}}$, which is $\sim3.5$ times more than in the inner region.

Based on the study of individual galaxies, we arrive at two major conclusions. First, the existence of excess 
X-ray sources appears to be a general phenomenon in the outskirts of galaxies. Second, the LMXB specific frequency 
is, on average, higher by a factor of $\sim3.5$ in the outer region than in the inner region, indicating that the 
excess X-ray sources do not originate from CXB fluctuations, and are not directly associated with the stellar 
populations in the outskirts of studied galaxies.

\subsection{The role of supernova kicks and GC-LMXBs}
\label{sec:subsample}

There are at least two plausible factors which could be responsible for the observed excess in the surface density of X-ray source excess in the galaxy outskirts. They could be LMXBs residing in globular clusters and/or accreting NS binaries that were kicked 
to large galactocentric radii by the supernova explosion. The 
importance of GC-LMXBs is higher in galaxies with large globular cluster specific frequency (\sn). On the other hand, one may expect that the effect of supernova kicks should depend on the mass of the galaxy. For example, high velocity binary systems may become unbound in case of low-mass galaxies and travel to very large distances, where they could blend with CXB sources. These possibilities are discussed in more detail in Sect.~\ref{sec:kicked}. 

Motivated by these considerations, we divided the sample into four different groups, each consisting of five 
galaxies. We used \sn$=2.0$ to separate galaxies with low/high \sn, and $M_{\rm{\star}}=1.2\times10^{11} \ \ M_{\odot}$ 
to distinguish low/high mass galaxies. The resulting four groups are: 
1) large \ms\ and large \sn, 2) large \ms\ and small \sn, 3) small \ms\ and large \sn, and 4) small \ms\ and small \sn. 
For each group we derived stacked radial profiles of the detected X-ray sources and computed the  predicted number of 
LMXBs as described in Sect.~\ref{sec:radial}. The obtained profiles are depicted in Fig. ~\ref{fig:groups}. A common property 
of the four groups is that in the inner region the observed profiles and profiles predicted based on the K-band light distribution and LMXB scaling relation are in fairly good agreement. 
However, in the outer region the first three groups show a significant source excess, with the only exception 
being low-mass galaxies with low \sn. Thus, we conclude that both globular cluster content and the mass of the galaxy are equally important factors in determining whether the excess at large galactocentric radii is present or not. As detailed in the following sections, our interpretation of this finding is that both GC-LMXB populations and supernova kicked systems contribute to the LMXB density enhancement in the galaxy outskirts.

\section{Combined XLFs of LMXBs}

\subsection{Accuracy of the CXB subtraction}
\label{sec:cxb}

To further explore the properties of LMXBs in the sample galaxies, we built combined XLFs of 
all the sources residing in the inner \inreg\ and outer \outreg\  regions. In total, we resolved $1412$ and $1077$ point sources in the inner and 
outer regions (Table~\ref{tab:stat}), allowing us to build  accurate XLFs. 

\begin{figure}
\resizebox{\hsize}{!}{\includegraphics[angle=0]{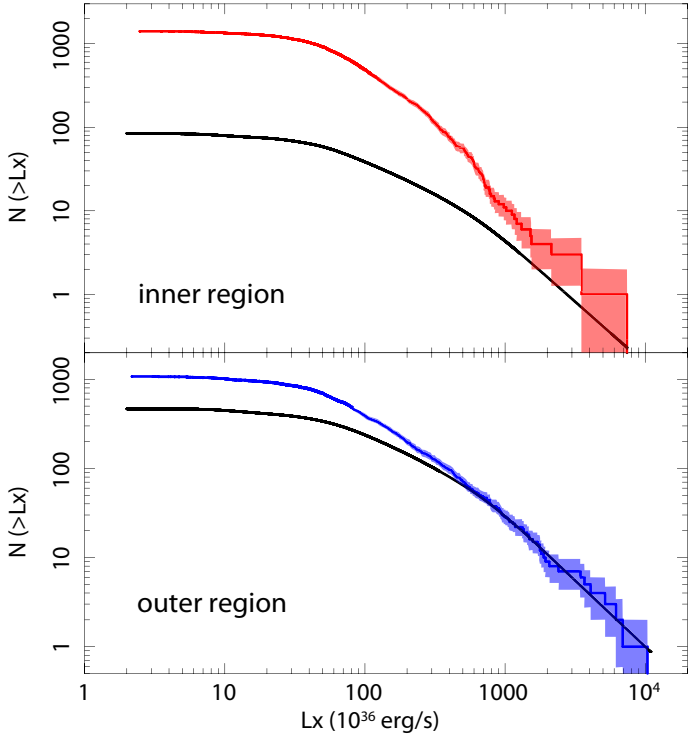}}
\caption{Cumulative luminosity distributions of all resolved point sources in the inner (upper) and outer 
(lower) regions of the sample galaxies. The distributions are not corrected for incompleteness or the contribution 
of CXB sources. The shaded areas correspond to 1$\sigma$ Poissonian uncertainties. The solid lines show the predicted 
 CXB luminosity distributions from \citet{Geo2008}, modified by the incompleteness function, as described in Sect. \ref{sec:chandra}.} 
\label{fig:cxb}
\end{figure}

Since our study is focused on galaxy outskirts with relatively low source density, it is crucial 
to accurately account for the CXB level. In the inner regions the estimated average contribution of CXB 
sources is only $\sim6\%$, whereas it is $\sim43\%$ in the outer regions (Table~\ref{tab:stat}). 
Given that the CXB source density exhibits $10-30\%$ field-to-field variations due to the cosmic 
variance, the inaccurate subtraction of the CXB level could significantly influence our study. Therefore, in Fig.~\ref{fig:cxb} we show the combined cumulative luminosity 
distributions of the detected X-ray sources in the inner and outer regions, which are compared with the 
predicted, incompleteness-corrected CXB source distributions. The upper panel of Fig.~\ref{fig:cxb} 
demonstrates that in the inner region the contribution of CXB sources is fairly low at luminosities 
below $10^{39} \ \rm{erg \ s^{-1}}$. Above this threshold we detected ultra-luminous X-ray sources, 
which are most likely stellar mass black holes accreting from a low or intermediate mass companion. 
For a comprehensive discussion of these ultra-luminous X-ray sources we refer to \citet{Zhang2012}.

The lower panel of Fig.~\ref{fig:cxb} shows that in the outer region the luminosity distribution of sources 
above $5\times10^{38}$ erg/s is in excellent agreement with the predicted CXB level. Indeed, above this luminosity  
 $71\pm8.4$ X-ray sources are detected, whereas 65.1 CXB sources are predicted. We note that in the bright end the XLF is 
not affected by incompleteness effects. Below $5\times10^{38}$ erg/s the number of detected sources is more than a factor 
of two of the expected number of CXB sources. Thus, in agreement with the  radial source density profiles (Fig.~\ref{fig:all_bright}), 
the XLF also shows the presence of an excess source population with luminosities $<5\times10^{38}$ erg/s
in the outer region. The good agreement between the observed source and predicted CXB luminosity distributions in the bright end demonstrates the accuracy of our CXB subtraction procedure. Moreover, the large difference between the XLFs in the faint end excludes the  possibility that the 
observed excess is due to the cosmic variance.

\begin{figure}
\resizebox{\hsize}{!}{\includegraphics[angle=0]{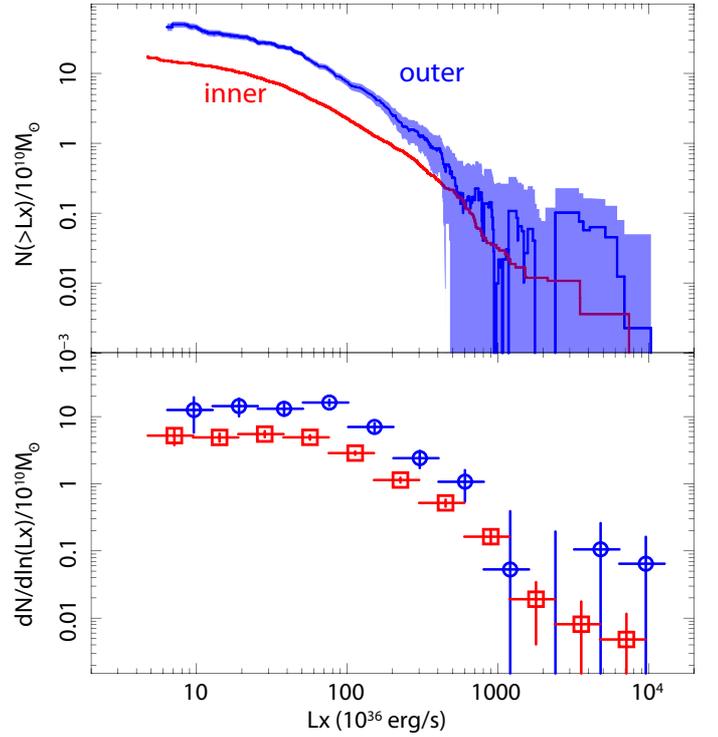}}
\caption{XLFs of LMXBs in the inner and the outer regions in cumulative (upper panel) and differential (lower panel) 
forms. The data for the outer regions is marked by circles in the lower panel and is surrounded by the shaded area 
showing the $1\sigma$ Poissonian uncertainty in the upper panel. The statistical uncertainties for the inner region have  smaller amplitude.} 
\label{fig:xlf}
\end{figure}

\subsection{Source XLFs of the inner and outer regions}
\label{sec:xlf}

We directly compare the combined XLFs of all galaxies in the inner and outer regions in Fig.~\ref{fig:xlf}. 
To build the XLFs, we only considered detected X-ray sources above the $0.6$ incompleteness level. 
The sources observed in individual galaxies were combined and weighted with the enclosed stellar mass 
within the regions following \citet{Zhang2011}. We subtracted the CXB level and applied incompleteness 
correction as described in Sect.~\ref{sec:chandra}. In the upper and lower panels of Fig.~\ref{fig:xlf} 
we show the final cumulative and differential forms of the XLFs, respectively.  

Although the statistical uncertainties at high luminosities are large, the XLFs of the inner and 
outer regions appear to be consistent with each other in the bright end. This conclusion is further supported 
by the luminosity distributions  shown in the Fig.~\ref{fig:cxb} and the shape of the radial profile of bright 
sources in the Fig.~\ref{fig:all_bright}. On the other hand, the outer region shows a significant source excess 
in the luminosity range below $\sim 5\times10^{38}$ erg/s. 
This threshold luminosity is fairly close to the Eddington luminosity of an accreting NS. Therefore, it is 
reasonable to assume that the bulk of faint excess sources are NS binaries. This conclusion is consistent with 
the proposed origin of excess sources. Indeed, both GC-LMXBs \citep{Zwart2000} and supernova kicked binaries 
\citep{Brandt1995} are predominantly NS binaries.

In principle, it would be beneficial to compare the XLFs of the four groups of galaxies with low/high \ms\ 
and \sn. The shape of the XLFs could help to constrain the origin of the excess sources, 
since field and GC-LMXBs exhibit markedly different XLFs at luminosities below $3\sim5\times10^{37}$ erg/s \citep{Zhang2011}. 
However, the studied sample is not suitable for such a comparison since the source detection sensitivity 
of most sample galaxies is comparable to this luminosity limit. Above $3\sim5\times10^{37}$ erg/s the XLFs 
of field and GC-LMXBs are virtually indistinguishable, hence we do not attempt a more detailed study of 
luminosity distributions.

\section{A case study for NGC 4365}
\label{sec:n4365}

The stacked source density profiles (Fig.~\ref{fig:groups}) indicate that the 
observed X-ray source excess 
in the galaxy outskirts cannot be explained only by the population of 
GC-LMXBs; it is very likely that supernova kicked 
accreting NSs also play a notable role. 
To separate the population of GC-LMXBs and supernova kicked sources, 
 GC-LMXBs must be identified at large galactocentric radii. However, for most galaxies in our sample 
this is not feasible because of the limitations of the available globular cluster data. 
Although several galaxies 
in our sample have been observed by the \textit{Hubble Space Telescope (HST)}, 
these images do not cover their
outskirts. Ground based observatories have a larger 
field-of-view, but these images usually 
suffer from incompleteness and foreground star contamination. An 
exception to this trend is NGC 4365, 
for which deep optical data is available, allowing us to detect GCs even in its 
outskirts. Additionally, 
deep \textit{Chandra} data is also available with a limiting luminosity of 
$\sim10^{37}$ erg/s. Therefore, we 
probe the X-ray populations of NGC 4365 as a case study. 

The galaxy NGC 4365 is a large elliptical galaxy with \ms $ =1.8\times10^{11} \ 
\rm{M_{\odot}}$ and \sn $=3.95$, which 
exhibits a notable source excess in the \outreg\ region. The 
GC population of NGC 4365 has been 
studied as part of the  SAGES Legacy Unifying Globulars and Galaxies Survey 
(SLUGGS). The galaxy and its outskirts 
have been observed in eight pointings with the Advanced Camera for Surveys (ACS) 
onboard \textit{HST}, whose data 
has been complemented with \textit{Subaru}/S-Cam observations. Based on this 
combined data set, \citet{Blom2012} 
reported more than $\sim$6000 GC candidates in and around NGC 4365. Beyond the 
central $0.5\arcmin$ ($\sim$1\re) 
the obtained list of GC candidates is complete to the turnover magnitude, 
and is not significantly affected by contamination \citep{Blom2012}. 

To build the list of GC candidates around NGC 4365, we relied on the publicly 
available SLUGGS\footnote{\textit{http://sluggs.swin.edu.au/data.html}} data set, 
combining the \textit{HST} and \textit{Subaru} 
identifications together. If a GC candidate was observed by both 
telescopes, we gave priority to \textit{HST}. 
We identified GC-LMXBs by cross-correlating the coordinates of GC candidates 
with the coordinates of 
detected X-ray sources following the method described in \citet{Zhang2011}. 
The applied match radius 
was $0.5^{\prime\prime}$, which results in a total number of 104 X-ray sources 
residing in GCs 
within the study field. The estimated number of random matches is $\sim3$. The 
104 identified sources 
are designated as GC-LMXBs, whereas all the others are field sources.

\begin{figure}
\resizebox{\hsize}{!}{\includegraphics[angle=90]{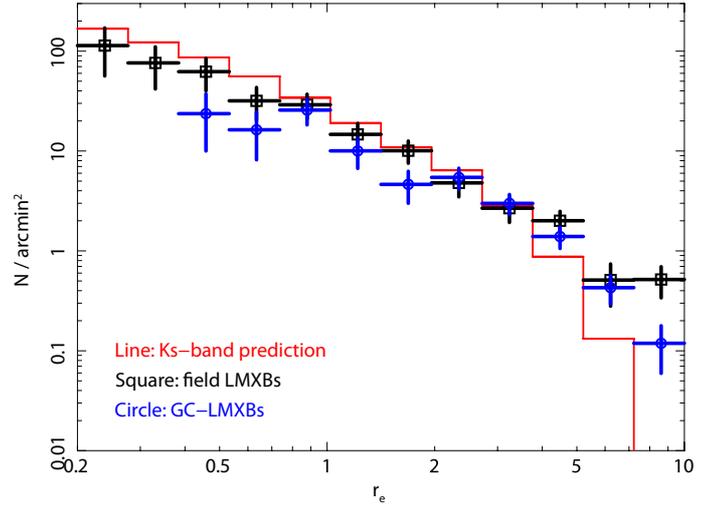}}
\caption{Stacked radial profiles of sources in NGC 4365. The 
CXB subtracted field LMXBs 
are marked with squares. The GC-LMXBs are marked with circles. The predicted 
distribution  of LMXBs 
from \ks-band light are plotted with histograms.} 
\label{fig:n4365}
\end{figure}

In Fig.~\ref{fig:n4365} we separately show the radial source density profiles 
of GC-LMXBs and 
 CXB subtracted field sources, which are compared with the expected 
number of LMXBs based 
on the enclosed stellar mass and the average LMXB XLF (Sect.~\ref{sec:stat}). 
In the \outreg\  
region 116 X-ray sources are detected (Table~\ref{tab:stat}). The predicted number of CXB sources for this region is 44.6, therefore there are $\sim 71.4$ compact sources associated with the galaxy. 
This is significantly larger than the expected number of LMXBs ($\sim 11.3$) predicted based on the K-band luminosity of this region and the standard LMXB scaling relation from \citet{Gilfanov2004}.
Among the detected sources, 25 sources are identified as GC-LMXBs, implying that 
there are $46.4\pm9.5$ field LMXBs, about four times the expected number. 

Thus, detailed study of  NGC 4365 provides a direct confirmation of the conclusion we made in Sect.~\ref{sec:subsample}, that the excess sources at large galactocentric radii are not all associated with  globular clusters.  In this particular case of a massive galaxy (even with relatively high \sn), almost $2/3$  of the excess sources must be of a different origin, for which, as we argue below,  supernova kicks seem to be a plausible scenario.

\section{Discussion}
\label{sec:discussion}

We have demonstrated that there are two equally important factors leading to the appearance of the excess LMXB populations in the outskirts of early-type galaxies. 
For a typical galaxy in our sample ($M_{\star} \sim 1.3\times10^{11} \ \rm{M_{\odot}}$ and $S_{\rm{N}} \sim 2.0$), about a half of the excess sources are due to LMXBs dynamically formed in blue globular clusters (having wider spatial distribution than stars). The origin of the other half cannot be explained by LMXBs residing in globular clusters. Instead, it is due to factors dependent on the mass of the galaxy. Below, we consider various possibilities and argue that a plausible scenario is that these are sources expelled to the outskirts of their parent galaxies as a result of the supernova kicks and we propose the mechanism, explaining the mass dependence of this effect.

\subsection{LMXBs associated with intracluster light}
\label{sec:icl}

In galaxy clusters the stellar light is not only associated with luminous
galaxies, but a certain fraction ($5-50\%$) of the total optical cluster
luminosity is associated with the intracluster light (ICL), which is bounded 
by the cluster potential \citep[e.g.,][]{Feldmeier2004,Gonzalez2005,Zibetti2005}. 
Thanks to its faint surface density, the ICL can only be detected with deep 
optical observations \citep{Mihos2005}, and remains essentially undetected in 2MASS 
images. Just as the stellar bodies of galaxies do, LMXBs can evolve in the ICL,
which in principle could contribute to the observed source excess in the
galaxy outskirts. Since part of the studied galaxies reside in a rich galaxy
cluster environment, we investigated the
importance of LMXBs associated with the ICL.

\begin{figure}
\resizebox{\hsize}{!}{\includegraphics[angle=0]{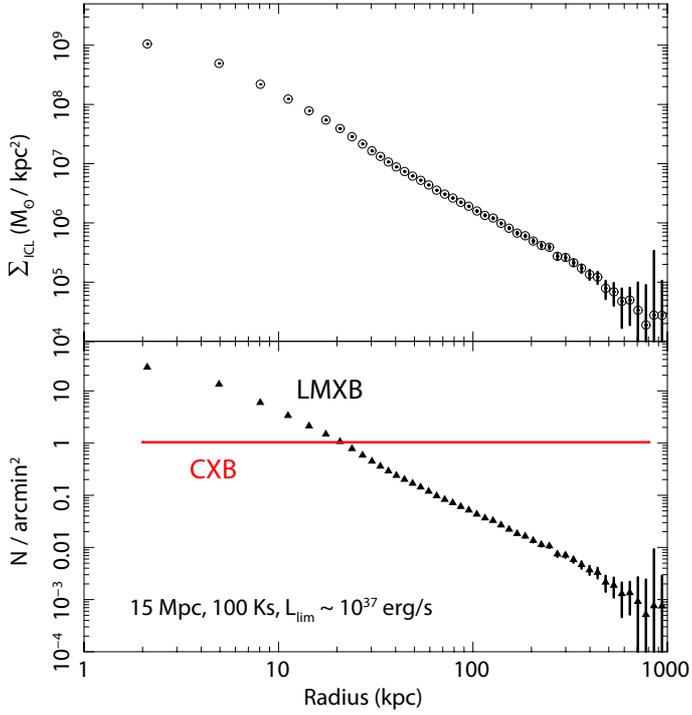}}
\caption{Upper panel: stellar mass density profile of the ICL calculated from 
\citet{Zibetti2005}. 
Lower panel: estimated number density of LMXBs associated with the ICL (triangles) 
and CXB sources (solid line), assuming a cluster at a distance of 15 Mpc and a detection sensitivity of 
$10^{37}$ erg/s.} 
\label{fig:icl}
\end{figure}

To estimate the average  surface density of the ICL, we relied on the
analysis of \citet{Zibetti2005}, who stacked the Sloan Digital Sky Survey
imaging data of 683 clusters in the redshift range of $z= 0.2 - 0.3$. We
used their stacked $I$-band ICL surface brightness distribution 
and assumed an $I$-band mass-to-light ratio of $M_{\star}/L_{\rm I}=1.50$ to
deduce the stellar mass. The average ICL mass density profile extending to
1000 kpc is shown in the upper panel of Fig.~\ref{fig:icl}. To convert the
stellar mass density profiles to the predicted LMXB source density profile, 
we assumed a galaxy cluster at a distance of 15 Mpc, adequate for
Virgo cluster, and a \chandra\ exposure time of 100 ks, which corresponds
to a detection sensitivity of $\sim10^{37}$ erg/s. Using these parameters
and the average ICL mass profile, we computed the number densities of the 
predicted LMXBs using their average XLF \citep{Zhang2012}. The obtained
source density profile reveals an LMXB density of $10^{-2}-10^{-3} \
\rm{arcmin^{-2}}$ in the radial range of $100-1000$ kpc (lower panel of
Fig.~\ref{fig:icl}). We note that several galaxies in our sample lie at even
larger radii from the cluster center, implying an even lower LMXB
density. Assuming the same distance and \chandra\ exposure time, we
estimated the CXB source density from their average log(N)--log(S) \citep{Geo2008}.
We obtained an average level of $\sim1 \ \rm{arcmin^{-2}}$, which is $2-3$
orders of magnitude higher than the expected LMXB number associated with
the ICL. Since the expected LMXB density from the ICL is only a minor
fraction of the CXB level, and the CXB level exhibits $10-30\%$
field-to-field variations, intracluster LMXBs are nearly impossible to detect and identify.
Thus, LMXBs associated  with the ICL will not add a notable contribution to the source
excesses detected in the galaxy outskirts.

\subsection{Stellar age dependence}
\label{sec:age}

\begin{figure}
\resizebox{\hsize}{!}{\includegraphics[angle=270]{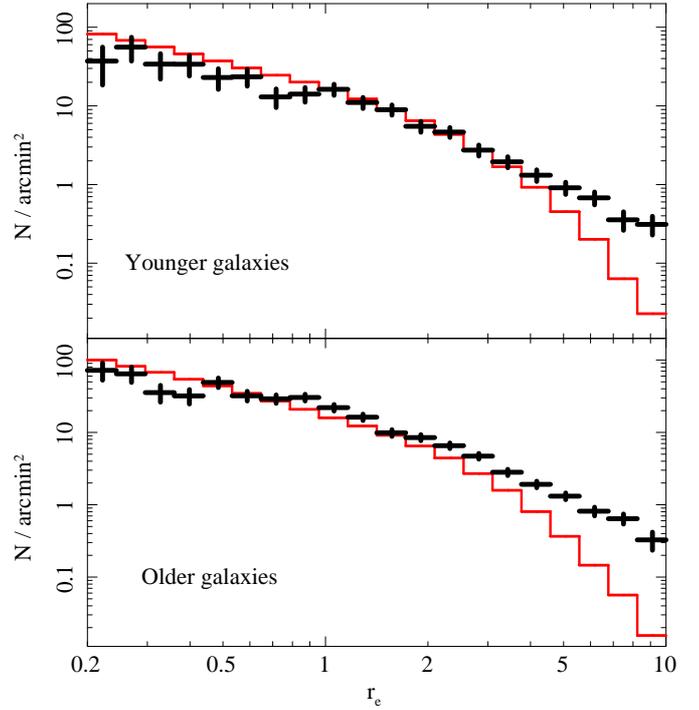}}
\caption{Stacked radial profiles of the  CXB subtracted LMXBs in the ten 
younger (upper panel)
and ten older (lower panel) galaxies. The estimated numbers 
of LMXBs from \ks-band 
light are plotted with solid histograms. The excess LMXB populations in the outer regions are
present for both younger 
and older galaxies.} 
\label{fig:young_old}
\end{figure}

In \citet{Zhang2012} we demonstrated that a correlation exists between the 
LMXB population and the stellar age of the host galaxy. Namely, the specific frequency of LMXBs is 
$\sim50\%$  higher in older galaxies than in younger ones. We demonstrated that this excess could be explained 
by the fact that older galaxies have larger GC-LMXBs populations, and by the intrinsic evolution of the LMXB population 
with time is likely to play a role. In the view of these findings we investigate whether the observed source excess 
in the outskirts of galaxies correlates with the stellar age. 

Following \citet{Zhang2012}, the sample of 20 galaxies was divided into 
younger and older galaxies using a median age of 6 Gyrs, which resulted in ten younger and ten 
older systems. We built stacked radial source density profiles for both sub-samples as 
described in Sect.~\ref{sec:radial}, and depicted them in Fig.~\ref{fig:young_old}. 
Compared to the predicted LMXB level, both young and old galaxies have a 
significant excess of X-ray sources in the \outreg\ region. In young galaxies 199.8 sources 
are observed above the CXB level in the outer region, which is two times larger than the predicted 
number of LMXBs from the  \ks-band light (66.8). For the old sub-sample we detect  409.6 sources 
above the CXB level, which is $3.5$ times more than the \ks-band prediction (91.3). 
Thus, the source excess appears to be more significant in older galaxies than in younger ones. 
This result can be interpreted as the result of two combined effects. First, a strong correlation 
exists between the stellar age and the specific frequency of GCs \citep{Zhang2012}. 
In our sample the median \sn\ is significantly higher for older galaxies (\sn 
$=4.64$) than for younger ones (\sn $=1.56$), implying that the frequency of GC-LMXBs 
is also notably higher for older galaxies in the outer regions.
Second, if younger galaxies indeed have smaller populations of primordially formed LMXBs, 
there would also be fewer kicked sources in the outskirts. We note that the median stellar 
mass of the young and old sample is comparable (1.53$\times 10^{10}\ 
\rm M_{\odot}$ and 1.47 $\times 10^{10}\ 
\rm M_{\odot}$, respectively), hence the importance of supernova kicks should be comparable.

However, because of the correlation between the stellar age and \sn, and the lack of
complete GC catalogs (Sect.~\ref{sec:n4365}), the population of GC-LMXBs
cannot be directly separated from supernova kicked neutron stars.
Therefore, based on the present sample, it is not feasible to
comprehensively study the frequency of supernova kicked sources as a
function of the stellar age.

\subsection{Quantifying the number of excess sources}
\label{sec:quantify}

To statistically probe the origin of excess X-ray sources in the \outreg\ 
region, 
we characterized the LMXB specific frequency in each galaxy using

\begin{eqnarray}
f_{\rm{XLF}} = \frac{N_{\rm{X}} - N_{\rm{CXB}} }{M_{\star} \times \int
F(L) K_{\rm{LMXB}} (L) dL} \ ,
\label{eq:fxlf}
\end{eqnarray}
where $N_{\rm{X}}$ and $N_{\rm{CXB}}$ are the numbers of resolved X-ray 
sources 
and the predicted CXB sources. For each galaxy, these values agree with those 
listed for the outer regions in Table~\ref{tab:stat}.
In Eq.~\ref{eq:fxlf} $M_{\star}$ is the stellar mass in units of $10^{10} \ \rm{M_{\odot}}$,
$F(L)$ is the average differential XLF normalized to $10^{10} \ 
\rm{M_{\odot}}$, 
and $K_{\rm{LMXB}} (L)$ is the incompleteness function for LMXBs in each 
galaxy. We note that this formula is essentially identical with that introduced by 
\citet{Zhang2012}.

According to Eq.~\ref{eq:fxlf}, if the number of X-ray sources agrees 
with that predicted from the average XLF, we expect to observe $f_{\rm{XLF}}=1$. 
However, for the studied 20 galaxies, we obtained the median value of $f_{\rm{XLF}}=3.31$ 
in the outer region, indicating a significant source excess. Assuming that the X-ray 
source excess depends on the GC frequency and mass of the galaxy, 
we fit the data with a two parameter linear model
\begin{eqnarray}
f_{\rm{XLF}} = a\times M_{\star} + b\times S_{\rm{N}} \ ,
\label{eq:fit}
\end{eqnarray}
where $M_{\star}$ is the stellar mass in units of $10^{10} \ \rm{M_{\odot}}$ 
and \sn\ is the globular cluster specific frequency. To find the best-fit values, 
we used $\chi^2$ minimization and obtained $a=0.081\pm0.025$ and $b=0.60\pm0.12$. 
The large $\chi^2/d.o.f=104.9/18$ implies notable scatter in the relation. 
Taking into account the median stellar mass ($M_{\star} = 1.3\times10^{11} \ 
\rm{M_{\odot}}$) 
and the median specific GC frequency ($S_{\rm{N}} = 2.0$) of our sample, we 
find that 
(on average) the contribution of kicked binaries ($a\times M_{\star} $) and 
GC-LMXBs 
($b\times S_{\rm{N}} $) is comparable.

\subsection{The role of supernova kicks}
\label{sec:kicked}

The fate of a kicked binary depends on the amplitude of the kick velocity  with respect to the orbital velocity in the binary system ($v_{\rm orb}$) and the escape velocity in the gravitational potential of the parent galaxy ($v_{\rm{esc}}$). If the final velocity attained by the binary system ($v_{\rm{sys}}$) exceeds the  escape velocity, the binary will become unbound from its parent galaxy. On the other hand, if the kick velocity is too large with respect to the orbital velocity  in  the binary system, it will be destroyed by the kick \citep[e.g.,][]{Brandt1995}. Because of this selection effect, the average system velocities of LMXBs are expected to  be smaller than for isolated neutron stars. 

\citet{Brandt1995} studied the effect of the supernova kick on the binary system. Assuming that the  birth velocities of neutron stars are distributed according to \citet{Lyne1994} with the mean value of 450 km/s, they concluded that LMXBs that remain bound after the supernova explosion have an average system velocity of  $v_{\rm{sys}}=180\pm80 \ \rm{km \ s^{-1}}$. The typical escape velocities of our sample galaxies are in the range of $250-1000 \ \rm{km \ s^{-1}}$ \citep{Scott2009}, larger than the average system velocity, therefore we should not expect a significant fraction of LMXBs to escape the parent galaxy, especially the more massive ones. Nevertheless, the system velocities are comparable to the escape velocities, suggesting that natal kicks of the compact objects can significantly modify the spatial distribution of LMXBs in the galaxy, leading to its broadening compared to the distribution of stars  \citep{Brandt1995}. However, this does not explain the observed dependence on the stellar mass. Assuming that the stellar mass is a proxy to the total gravitating mass, we should expect that the effect of kicks is stronger in the lower mass galaxies, the opposite of what is observed. 

It has been suggested, however, that the birth velocity distribution of the Galactic radio pulsar population has a bimodal structure, with three-dimensional dispersions of the Gaussian components $\sim 90$ km/s and $\sim 500$ km/s and the population roughly equally divided into the two components \citep{Arzoumanian2002}. The bimodal distribution of the natal kicks will result in the similar bimodal velocity distribution of LMXBs \citep[e.g.,][]{Repetto2012}. Thus, one may expect that the lower velocity component remains bound in galaxies of any mass from our sample. The fate of the high velocity component, however, depends on the gravitational mass of the galaxy.  In the higher mass galaxies it will remain bound but would be much more widely distributed than stars, because of the much larger average velocity, thus giving rise to the excess of X-ray sources in the galactic outskirts. In the case of the lower mass galaxies the majority of the high velocity systems will escape the galaxy and no extended halo of LMXBs will appear.
            
The escaping binaries will travel to large distances in a relatively short time scale. Assuming a system velocity of $v_{\rm{sys}}=200 \ \rm{km \ s^{-1}}$ and a radial trajectory, we estimate that in $1$ Gyr the binary will move from the host galaxy to a distance of $\sim200$ kpc. This value significantly exceeds the field-of-view of the presently studied \chandra\ observations ($30-60$ kpc), implying that such kicked X-ray sources cannot be detected in our sample. On another note, owing to the expected low surface density of supernova kicked sources at large radii, they will be virtually indistinguishable from the CXB level.

As the exact cause of the natal kick is not understood, it is not known what the birth velocities of black holes are. Unlike neutron stars, they cannot be measured directly from proper motions of isolated black holes, while attempts to derive them from the spatial distribution of black hole binaries so far has produced ambiguous results. Among different models, a momentum-conserving model is often considered, which assumes that during the supernova explosion, black holes and neutron stars receive the same momentum (rather than the same kick velocity). In this model, the kick velocities of black holes are smaller by a factor of $5-10$ than those received by the neutron stars, according to their mass ratios. Therefore black hole kicks would be insufficient to drive them out of even the lowest mass galaxies in our sample. This agrees with the fact that the threshold luminosity above which the extended halo of X-ray sources vanishes is about $(3-5)\cdot 10^{37}$ erg/s, which is close to the Eddington luminosity limit for the neutron star. It also suggests that the excess sources in the galactic outskirts with supernova kicked origins are neutron star binaries.
        
Above, we outlined a qualitative picture which may become a framework for a quantitative analysis. A possible direct proof of the kick scenario could be a measurement of the line-of-site velocities of X-ray sources located at large galactocentric radii, which may be a daunting (but still possible) task given the optical faintness of these systems. On the other hand, this picture can receive further support or it can be falsified with realistic calculations of the dynamics of the kicks and kinematics of the kicked systems similar to the ones performed by \citet{Brandt1995} and \citet{Repetto2012}. Results of such calculations can be compared with the observed radial source density profiles in galaxies of different masses. A comparison of this kind would also require detailed globular cluster information in order to remove the contribution of globular cluster LMXBs. This, however, is beyond the scope of this paper.

\section{Conclusion}
\label{sec:conclusion}

In this paper, for the first time, we have systematically explored the
population of LMXBs in the outskirts of early-type galaxies. We have
studied a sample of 20 early-type galaxies based on archival \chandra\
data which allowed us to perform a statistically significant study of the
X-ray populations with a limiting luminosity of $\sim 10^{37} \ \rm{erg \
s^{-1}}$. Our results can be summarized as follows.
\begin{enumerate}
\item We demonstrated the existence of an excess X-ray source population
in the outskirts of early-type galaxies. These sources form a halo of compact X-ray sources around galaxies extending out to at least $\sim 10r_e$ and probably much further. Their radial distribution is much wider than the distribution of the stellar light. The extended halo is comprised of sources with luminosity $\la 5\cdot 10^{38}$ erg/s, while the more luminous sources appear to follow the distribution of the stellar light without any notable excess at large radii.  This suggests that the majority of the excess sources are neutron star binaries.
\item Dividing the sample galaxies into four different groups based on
their stellar mass and the specific frequency of globular clusters, we found that the extended halo of compact sources  is present in all groups except for the low-mass galaxies with low globular cluster content (small $S_{\rm{N}}$). 
\item We performed a case study of NGC 4365, for which deep optical
(\textit{HST} and \textit{Subaru}) and X-ray data (\chandra) is available,
allowing the identification of GC-LMXBs. In the \outreg\ region we detected
$60.1\pm10.8$ excess X-ray sources, out of which 25 are GC-LMXBs and
$35.1\pm9.5$ are field sources which are presumably supernova kicked 
NS binaries.
\item Interpreting our findings, we propose that the excess sources are comprised of two independent components: (i) dynamically formed sources in blue (metal poor) globular clusters, which are known to be distributed wider than the stellar light and; (ii) primordial (field) neutron star LMXBs expelled from the main body of the galaxy as a result of kicks received during the supernova explosion. 
\end{enumerate}

\begin{acknowledgements} 

We thank Andrew Cooper, Diederik Kruijssen, and Jingying Wang for discussions
which have improved this paper. We thank Klaus Dolag for the discussion of the intracluster 
light in clusters of galaxies. This research has made use of \chandra\ archival data provided by
\chandra\ X-ray Center, and 2MASS Large Galaxy Atlas data provided by
NASA/IPAC infrared science archive. {\'A}kos Bogd{\'a}n acknowledges 
support provided by NASA through Einstein Postdoctoral Fellowship 
grant number PF1-120081 awarded by the Chandra X-ray Center, which 
is operated by the Smithsonian Astrophysical Observatory for NASA
under contract NAS8-03060. 

\end{acknowledgements}

\bibliographystyle{aa}
\bibliography{ms}

\end{document}